\documentclass[review]{elsarticle}

\usepackage{amsmath} 

\usepackage{lineno,hyperref}

\usepackage{url}  
\usepackage{graphicx} 
\usepackage{bm}
\usepackage{textcomp}
\usepackage{revsymb}
\usepackage{amssymb}
\usepackage{ifsym}
\usepackage[nointegrals]{wasysym}
\usepackage{txfonts}
\usepackage{tipa}

\begin{document}
	
	\begin{frontmatter}






		
		\title{Spectral properties of orthorhombic {KSbOSiO$_4$} crystals: studies of infrared, visible and ultraviolet radiation absorption, Raman scattering, cathodoluminescence and photoluminescence 
		}


\author[GPI]{Vladimir~A.~Yuryev\corref{Yuryev}}
\cortext[Yuryev]{Corresponding author}
\ead{vyuryev@kapella.gpi.ru}
\author[GOSNIIR]{Tatyana~V.~Yuryeva} 
\ead{yuryevatv@gosniir.ru}
\author[GOSNIIR]{Ekaterina~A.~Morozova}
\ead{morozovaea@gosniir.ru}
\author[GOSNIIR]{Irina~F.~Kadikova} 
\ead{kadikovaif@gosniir.ru}
\author[INME]{Polina~A.~Edelbekova}
\ead{edelbekova.p@inme-rus.ru}

\address[GPI]{A.\,M.\,Prokhorov General Physics Institute of the Russian Academy of Sciences,\\38 Vavilov Street, Moscow 119991, Russia}

\address[GOSNIIR]{The State Research Institute for Restoration, Building 1, 44 Gastello Street, Moscow 107114, Russia}


\address[INME]{Institute of Nanotechnology of Microelectronics of the Russian Academy of Sciences, 32A~Leninsky Avenue, Moscow 119991, Russia}

\cortext[ORCID]{ORCID~IDs: 0000-0001-8236-0728 (Vladimir~A.~Yuryev), 0000-0003-1378-330X (Tatyana~V.~Yuryeva), 0000-0002-7042-0997 (Ekaterina~A.~Morozova), 0000-0002-7271-0752 (Irina~F.~Kadikova), 0000-0002-8422-9798 (Polina~A.~Edelbekova)}



\begin{abstract}
Crystals of orthorhombic KSbOSiO$_4$ (KSS) are often found to synthesize in glass of historical and artistic articles.
For instance, these crystallites in the glass of turquoise Bohemian seed beads are known to play a significant role in the process of internal-stress driven corrosion of the glass causing deterioration and destruction of bead embroidered museum exhibits, especially articles made in the early 19th century.
In art materials science, KSS has been known since the mid-2010s only for its destructive effects on historical glass objects, but as an optical material it has long been studied.
KSS belongs to the group of compounds isomorphic to KTiOPO$_4$ (KTP) and, like other compounds from this group, has anomalies in dielectric constant.
The article presents the results of a study of orthorhombic KSS crystals using a number of spectroscopic techniques.
IR absorption and diffuse reflection spectra of KSS in the spectral range from 50 to 4000~cm$^{-1}$,
its Raman spectrum,
its absorption spectrum in the range from near infrared to ultraviolet
as well as
its cathodoluminescence and photoluminescence spectra
are investigated.  
The constituent peaks of the spectral bands are analyzed and their properties are summarized in tables;
the allowed direct and indirect band gaps of KSS are determined.
The presented data are necessary both for the correct spectral analysis of historical and artistic glass and enamels, during the production of which this substance was sometimes synthesized in glass, and for a more in-depth study of crystals of the KTP group.
\end{abstract}

\begin{graphicalabstract}	
	\includegraphics[width=\textwidth]{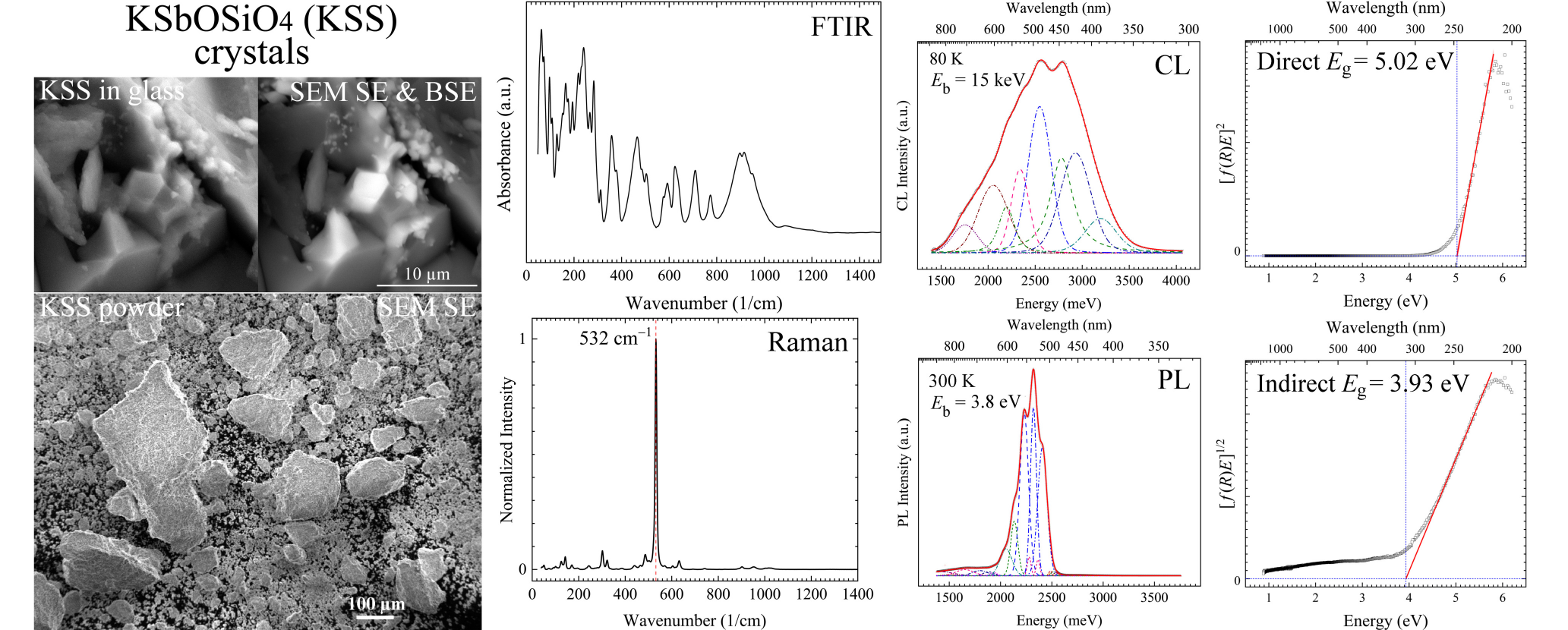}
\end{graphicalabstract}

\begin{highlights}
	\item 
	FTIR and Raman spectra of orthorhombic KSbOSiO$_4$ powder are analyzed in detail
	\item 
	KSS absorption and reflection spectra in the range from near IR to UV are studied
	\item 
	Cathodoluminescence and photoluminescence spectra of orthorhombic KSS are studied
	\item 
	Constituent peaks of KSS spectral bands are analyzed, their parameters are tabulated 
	\item 
	The direct and indirect band gaps of KSS are assessed using the Kubelka-Munk model
\end{highlights}

\begin{keyword}
\texttt{potassium antimony silicate \sep IR absorption \sep Raman scattering \sep cathodoluminescence \sep photoluminescence \sep visible$-$ultraviolet absorption}
\end{keyword}

\end{frontmatter}


\newpage

\section{Introduction}
\label{sec:intro}

\begin{figure}[ht]
	\includegraphics*[width=\textwidth]{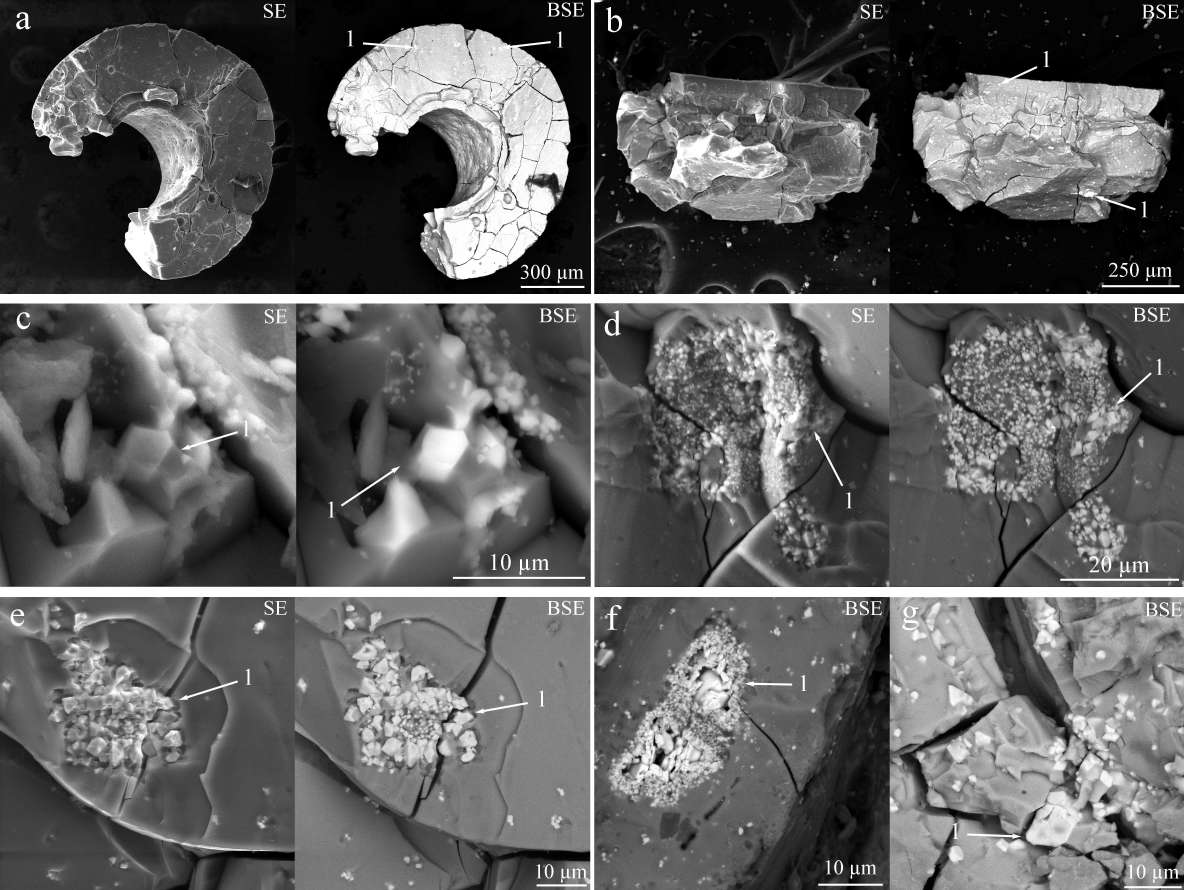}%
	\caption{SEM SE and BSE images of glass particles from unstable 19th century turquoise glass beads (a, b) fragmented due to internal corrosion, and KSS crystals in the glass of these beads (c--g); 
		figures 1 and arrows indicate KSS colonies in the bead glass; 
		the KSS colonies are seen to cause glass fracture due to mechanical stress built in the glass during its cooling after tumble finishing of the beads \cite{KSS_Electron_microscopy, 2020_Glas_depolimer, 2022_Glas_alter}
		\label{fig:Glass_SEM}	
		}
		       
\end{figure}

It has been reported in a number of works that microcrystallites of orthorhombic KSbOSiO$_4$ (KSS) are largely responsible for the long-term corrosion and destruction of turquoise seed beads of the 19th century made of lead-potassium glass (Fig.~\ref{fig:Glass_SEM}) leading to irreversible deterioration of bead embroidered exhibits kept in museums \cite{Yuryev_JOPT, Yur_JAP, Beads_KSS_SPIE, KSS_Electron_microscopy}.
It was found that these crystals and their colonies had synthesized in the glass during its melting \cite{KSS_Electron_microscopy}. 
Then, when glass tubes \cite{KSS_Electron_microscopy, Yurova_large, Glass_Bead-Making}, blanks for the manufacture of seed beads, or the cut beads cooled down after high-temperature processes such as tube forming, thinning and especially bead tumble finishing at elevated temperature, due to the difference in the linear expansion coefficients of KSS and glass, internal stress fields arose around the crystallites, which, in the course of time, resulted in the formation of micro discontinuities and then micro cracks in the glass \cite{KSS_Electron_microscopy}.
The latter turn into large fissures in decades that eventually causes bead crumbling into sand-like particles.
The thermal-fluctuation mechanism explaining the crack formation process was considered in our previous works \cite{2020_Glas_depolimer, 2022_Glas_alter}.
This mechanism also accounts for other slow moving processes leading to glass fracturing and decay, namely depolymerization of glass and the migration of alkali metal atoms along the stress gradient \cite{2022_Glas_alter}.%
\footnote{%
	The phenomenon of directed migration of impurities in glass caused by internal stress is analogous to the diffusion of metal atoms that occurs during the process of internal gettering \cite{Internal_gettering-1,Internal_gettering-Multy-Si,Internal_gettering-Si-PV}, which is often employed for the integrated circuit manufacturing and also can be described in terms of the thermal-fluctuation mechanism. 
} 

Thus, KSS crystallites embedded in glass can be a source of strong internal stress in the glass, which often leads to glass destruction and irreversible deterioration of museum exhibits.
In particular, in the case of beaded items, this leads to the need to replace corroded or crumbling antique beads with modern ones that are noticeably different from the historical ones in both shades and sizes. 
This necessitates the detection of KSS in glass of museum items using non-invasive optical spectroscopy techniques and, therefore, generates interest in its spectral properties.

It is known from literature that, in addition to the above-mentioned Bohemian turquoise beads of the 19th century, KSS has also been found in some other glass elements of archaeological, historical and artistic objects, e.g., in the blue tesserae of the mosaics decorating the \textit{Grotta del Buontalenti} (Giardino di Boboli, Florence, Italy) \cite{KSS_in_tesserae} or in some turquoise glass trade beads excavated at archaeological sites in southern Africa and analyzed by Raman spectroscopy \cite{Lazurite_error-arXiv, Lazurite_error, Raman_KSS_Brandt}.


In art materials science, KSS has been known since the mid-2010s only for its destructive effects on historical glass objects, but it has been studied as an optical material for a long time.
The fact is that KSS belongs to the group of compounds isomorphic to KTiOPO$_4$ (KTP) \cite{35A-2_KSbOSiO4, COD-KSbOSiO4, KSbOSiO4_synth+struct, KTP-KSS_sol-gel} and, like other substances from this group, has anomalies in dielectric constant, yet exhibiting relaxation behavior \cite{35A-2_KSbOSiO4, KSS-ferroelectric, KSS-dielectric_relax, KSS-nonlinear}.
KTP crystals are known to be highly ferroelectric with a high Curie temperature ($T_{\mathrm{C}} \approx 935$\textcelsius).
As Ti and P are replaced by Sb and Si the transition to KSS occurs, the Curie point drops considerably.
Pure KSS exhibits a phase transition at significantly lower temperatures \cite{35A-2_KSbOSiO4, KSS-ferroelectric, KSS-dielectric_relax}.
Solid solutions between KTP and KSS ({K(Ti$_{1-x}$Sb$_x$)O(P$_{1-x}$Si$_x$)O$_4$}) at certain values of $x$ begin to exhibit properties characteristic of relaxors \cite{KSS-ferroelectric, KSS-dielectric_relax}. 
The decrease in the nonlinear optical efficiency of KSS occurs because the Sb$^{5+}$ ions in the [SbO$_6$] octahedra are arranged much more symmetrically than the strongly shifted Ti$^{4+}$ ions in the KTP, as a consequence, 
the giant nonlinear response practically disappears in KSS, and the second harmonic generation efficiency of KSS is negligible compared to KTP.

Nevertheless, we believe that KSS crystals, like other KTP group crystals, are of interest for studying their spectral properties in order to gain deeper knowledge about KTP group crystals.
The Fourier transform infrared absorbance spectrum of the synthesized orthorhombic KSbOSiO$_4$ was shown for the first time in Ref.~\cite{KSbOSiO4_IR_absorption};
the knowledge of its characteristic bands helped us to analyze the silicate unit composition of turquoise bead glass in well preserved and corroded beads \cite{2020_Glas_depolimer} and especially in intact and degraded domains of beads \cite{2022_Glas_alter, Preservation_Cultural_Heritage-2018}.
The Raman spectrum of the synthesized orthorhombic KSbOSiO$_4$ powder was presented in Refs.\,\cite{Beads_KSS_SPIE, Raman_KSS_Brandt, KSS_Raman_spectrum}; 
its characteristic feature is an intense narrow peak at about 532~cm$^{-1}$.
The cathodoluminescence spectra of orthorhombic KSbOSiO$_4$ recorded at 80 and 300~K were demonstrated for the first time in Ref.\,\cite{KSS_Electron_microscopy}; 
their characteristic band has two maxima at about 444 and 483~nm at both temperatures.

In this article, characteristic Fourier transform infrared absorption spectrum of a specially synthesized crystalline powder of orthorhombic KSS is investigated in detail in the spectral range from 50 to 4000~cm$^{-1}$.
Its characteristic Raman spectrum as well as its absorption and diffuse reflection spectra in the range from near infrared to ultraviolet, and its cathodoluminescence and photoluminescence spectra are also studied. 
Constituent peaks of the detected spectral bands are thoroughly analyzed and their parameters are tabulated;
the widths of allowed direct and allowed indirect band gaps of orthorhombic KSS are determined from the diffuse reflection spectrum using the Kubelka-Munk theory.



\section{Sample preparation, experimental methods and equipment} 
\label{sec:experiment}

\subsection{Samples}
\label{sec:samples}

\begin{figure}[t]
	\includegraphics*[width=\textwidth]{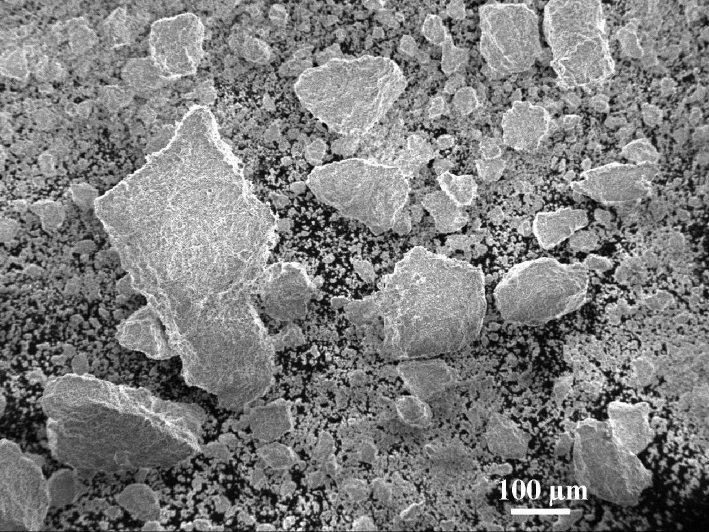}%
	\caption{SEM SE image of the synthesized orthorhombic KSbOSiO$_4$ powder
		\label{fig:KSS_SEM}	 
	}
	\end{figure}

Samples of orthorhombic KSbOSiO$_4$ were synthesized following the process described in Ref.\,\cite{35A-2_KSbOSiO4, COD-KSbOSiO4}: 
KNO$_3$, Sb$_2$O$_3$ and SiO$_2$ taken in stoichiometric proportion were annealed at the temperature of 1100{\textcelsius} in air ambient for 24 hours. 
The resultant samples were powder of polycrystalline particles of various sizes (Fig.~\ref{fig:KSS_SEM}).

X-ray powder diffraction pattern of the obtained compound proved that a single phase of orthorhombic KSbOSiO$_4$ \cite{35A-2_KSbOSiO4} was synthesized (space group \textit{Pna}2$_1$ (33); CAS No.~132265-14-6; PDF~No.~00-045-0323) \cite{Yur_JAP, Beads_KSS_SPIE}. 
Peaks associated with the tetragonal phase \cite{COD-KSbOSiO4,Tetragonal-orthorhombic} were not observed in the powder patterns as well as peaks, which could be attributed to other substances.%
\footnote{%
Samples of orthorhombic KSS for Raman spectroscopy studies carried out in Refs.~\cite{Beads_KSS_SPIE, Raman_KSS_Brandt, KSS_Raman_spectrum} were synthesized by E.~A.~Morozova in the same batch as the samples studied in this work.}

\subsection{Methods and instruments}
\label{sec:methods}

\paragraph{Scanning electron microscopy}
Scanning electron microscopy (SEM) images of KSS crystallites in glass given in Fig.~\ref{fig:Glass_SEM} were obtained using Vega-II~XMU, Mira~3~LMH or Mira~3~XMU microscopes (Tescan Orsay Holding). 
As a rule, SEM images were recorded in both backscattered electron (BSE) and secondary electron (SE) modes since SEM operating in the former mode represents mainly the substance elemental composition (i.e. its density), while operating in the latter mode it renders mainly the spatial relief, whereas both types of the information are usually required to correctly interpret data obtained using SEM.
An image of the KSS powder in Fig.~\ref{fig:KSS_SEM} was obtained using JSM-7001F SEM (Jeol).

\paragraph{Fourier transform infrared spectroscopy}
Fourier transform infrared (FTIR) spectra of KSS samples in the spectral range from 600 to 4000~cm$^{-1}$  were obtained using a LUMOS microscope (Bruker) in attenuated total reflection (ATR) mode with a Ge ATR crystal.
FTIR spectra recorded in the interval between 50 and 4000~cm$^{-1}$ were obtained using a Vertex 70v spectrometer (Bruker) in ATR mode with a diamond ATR crystal.  
Experiments were carried out at a spectral resolution of 4~cm$^{-1}$; routinely, 64 scans were made.

\paragraph{Raman spectroscopy} 
Raman spectra of KSS were recorded at room temperature with an InVia Qontor confocal microspectrometer (Renishaw) 
using an RL532C100 (Renishaw)
laser 
with the wavelength $\lambda_{\mathrm{ex}}=532$~nm and maximum emission power $P_{\mathrm{max}}=100$~mW.
The diffraction grating with 1200 gr/mm was used. 
The lens magnification was 50$\times$.

\paragraph{Spectroscopy of UV, visible and near IR radiation absorption and reflection}
Spectra of absorption (LA) and diffuse reflection (LR) of UV, visible and near IR radiation were investigated in the spectral range from 200 to 1400~nm using an UV-2600 spectrophotometer (Shimadzu);
the instrument was equipped with an IRS-2600Plus integrating sphere.

\paragraph{Microcathodoluminescence}
Cathodoluminescence (CL) spectra of KSS were registered both at room tem\-perature and liquid nitrogen temperature using a Mono\-CL3 system (Gatan) mo\-un\-ted at JSM-7001F SEM (Jeol).
A Peltier-cooled photomultiplier tube for the spectral range from 160 to 930 nm and a set of blaze gratings were used for spectra acquisition. 

\paragraph{Microphotoluminescence} 
Photoluminescence (PL) spectra of KSS were recorded at room temperature with an InVia Qontor confocal microspectrometer (Renishaw) 
using an IK3301R-G He--Cd laser (Kimmon Laser Systems) with the wavelength $\lambda_{\mathrm{ex}}=325$~nm and maximum emission power $P_{\mathrm{max}}=35$~mW.
The diffraction grating with 1200 gr/mm was used. 
The lens magnification was 15$\times$.

\section{Results: spectra of orthorhombic KSbOSiO$_4$ and their analysis} 
\label{sec:results}

\subsection{Absorption and Raman bands of orthorhombic KSbOSiO$_4$ in the spectral range from 50 to 4000~cm$^{-1}$ and 
	their constituent peaks }
\label{sec:FTIR}

\begin{figure}[t]
	\includegraphics[width=\textwidth]{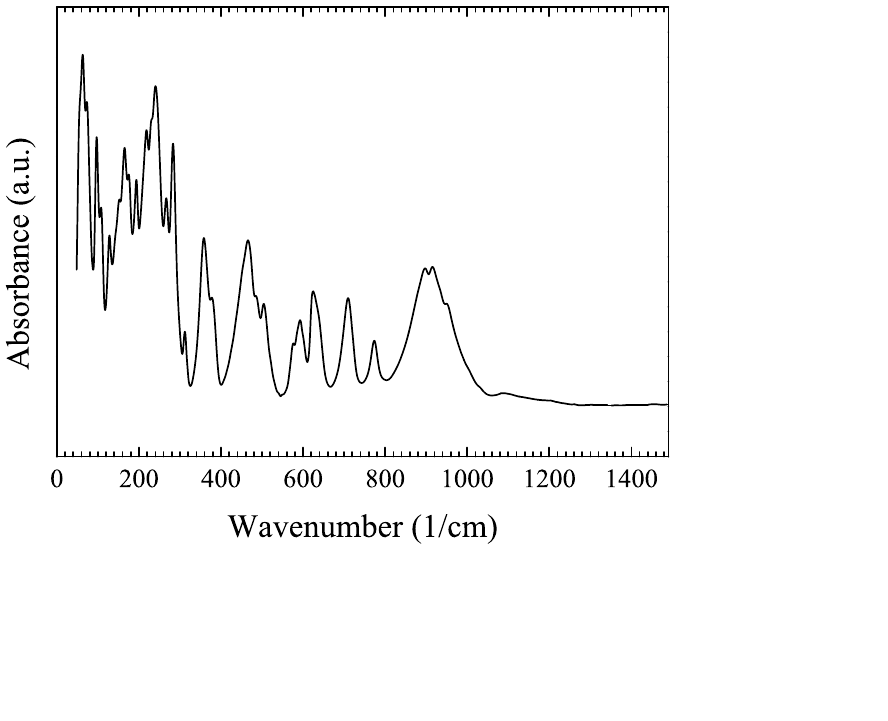}%
	\caption{A survey FTIR absorbance spectrum of orthorhombic KSbOSiO$_4$ \cite{KSbOSiO4_IR_absorption}
		\label{fig:survey}	 
	}
\end{figure}

Absorption spectrum of orthorhombic KSS powder (Fig.~\ref{fig:survey}) demonstrates a rich set of spectral bands both in the mid-IR range from 325 to 1300~cm$^{-1}$ (we designate this range with the letter M) and in the far-IR range from 40 to 325~cm$^{-1}$ (we designate it with the letter F).

Survey Raman spectrum of orthorhombic KSS \cite{KSS_Raman_spectrum} is shown in Fig.~S1\,a (\ref{ESM1}).
The most intense line in the spectrum peaks at a Raman shift of 532~cm$^{-1}$.
This is the characteristic line for detection of orthorhombic KSS in Raman analysis \cite{Beads_KSS_SPIE, Raman_KSS_Brandt}.

\subsubsection{Mid-IR absorption bands}
\label{sec:MIR}

\begin{figure}[t]
	\includegraphics[width=\textwidth]{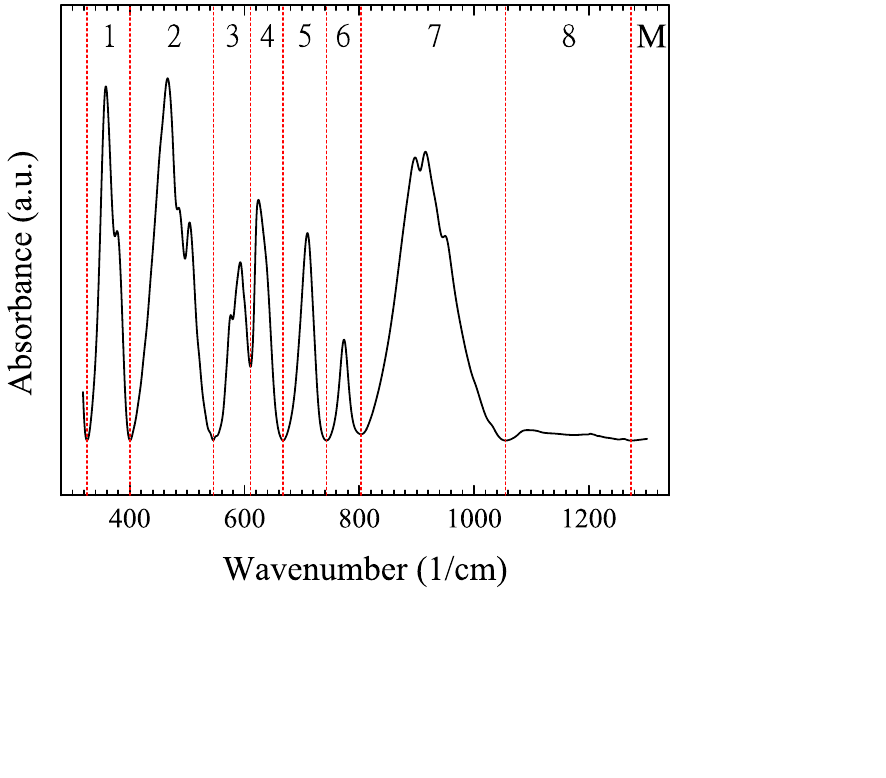}
	\caption{Mid-IR absorbance spectrum of orthorhombic KSbOSiO$_4$; characteristic spectral bands are marked with numerals from 1 to 8, letter M in band designation means that it belongs to mid-infrared spectral range
		\label{fig:MIR}	 
	}
\end{figure}

Fig.~\ref{fig:MIR} demonstrates the normalized IR spectrum of KSS in the M range with the baseline subtracted.
We have detected 8 spectral bands in this region, numbered  M\,1 to M\,8 in Fig.~\ref{fig:MIR}.
Their parameters and results of their peak analysis are presented in Table~\ref{tab:Mid-IR_bands}.

\begin{figure}[t]
	\includegraphics*[width=\linewidth]{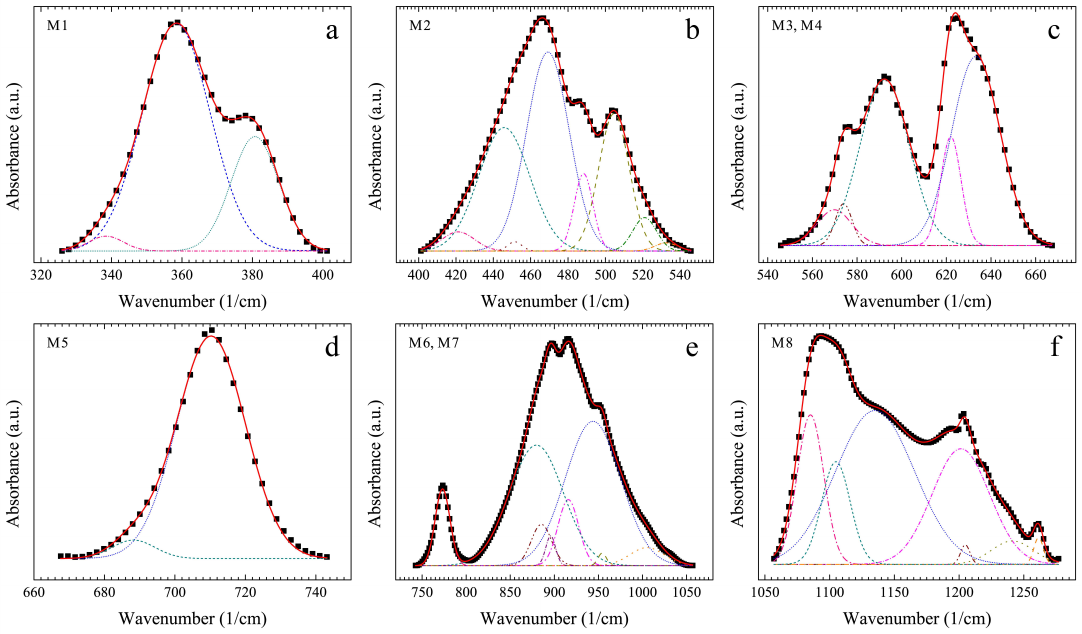}
	\caption{Absorption bands of orthorhombic KSbOSiO$_4$ in mid-IR range and 
their deconvolution using
Voigt functions 
		\label{fig:MIR_bands}	 
	}
\end{figure}

Band M\,1 reaching maximum at a wavenumber of 358~cm$^{-1}$ contains 2 local maxima at 358 and 378~cm$^{-1}$ (Fig.~\ref{fig:MIR_bands}\,a); 
it is composed by 3 lines peaking at 338, 358 and 381~cm$^{-1}$.
Deconvolution of the following band M\,2 (Fig.~\ref{fig:MIR_bands}\,b) with a maximum at 466~cm$^{-1}$ and 3 local peaks at 466, 485 and 504~cm$^{-1}$ yields 8 constituents at 
421, 446, 451, 469, 488, 505, 521 and 533~cm$^{-1}$. 
%
The next two bands M\,3 and M\,4 (Fig.~\ref{fig:MIR_bands}\,c) with maxima at 593 and 624~cm$^{-1}$ and local peaks at 576 and 593~cm$^{-1}$ (M\,3) and 624~cm$^{-1}$ (M\,4) are decomposed into 5 and 3 spectral lines, respectively;  
their maxima are at 550, 569, 575, 592 and 603~cm$^{-1}$ (M\,3), and 621, 628 and 634~cm$^{-1}$ (M\,4). 
Band M\,5 has a single maximum at 710~cm$^{-1}$ (Fig.~\ref{fig:MIR_bands}\,d);
it is composed by 2 lines centered at 691 and 710~cm$^{-1}$.
Band M\,6 consisting of one spectral line peaks at 773~cm$^{-1}$ (Fig.~\ref{fig:MIR_bands}\,e);
band M\,7 with a maximum at 915~cm$^{-1}$, plotted on the same panel, demonstrates 3 local peaks at 897, 915 and 950~cm$^{-1}$;
its deconvolution yields 9 lines peaking at  
879, 886, 
897, 916, 
933, 945,
954, 1006
and
1033~cm$^{-1}$.
Finally, peak analysis of a weak band M\,8 with a maximum at 1093~cm$^{-1}$ and 4 local peaks at wavenumbers of 1093, 1194, 1203 and 1261~cm$^{-1}$ gives 9 components peaking at
1085, 1105, 
1133, 1200,
1203, 1209, 
1221, 1237
and
1261~cm$^{-1}$ (Fig.~\ref{fig:MIR_bands}\,f).


\subsubsection{Far-IR absorption bands}
\label{sec:FIR}

\begin{figure}[t]
	\includegraphics*[width=\linewidth]{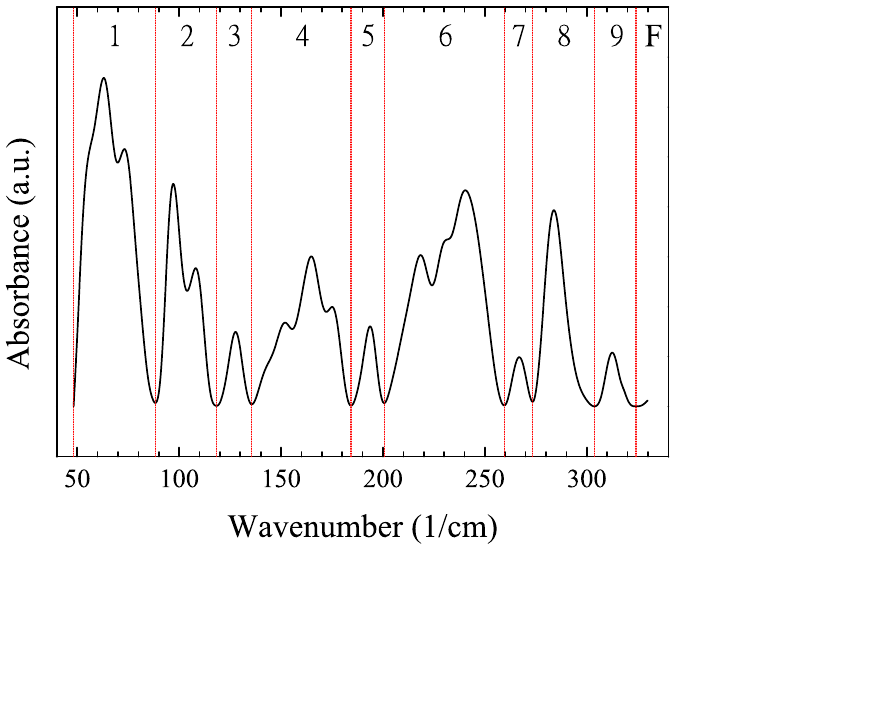}
	\caption{Far-IR absorbance spectrum of orthorhombic KSbOSiO$_4$; characteristic spectral bands are marked with numerals from 1 to 9, letter F in band designation means that it belongs to far-infrared spectral range
		\label{fig:FIR}	 
	}
\end{figure}

Fig.~\ref{fig:FIR} presents the normalized IR spectrum of KSS in the F range after subtraction of the baseline.
There are 9 spectral bands allocated in this range, numbered by us from F\,1 to F\,9.
Their parameters and results of their peak analysis are gathered in Table~\ref{tab:Far-IR_bands}.

\begin{figure}[ht]
	\includegraphics*[width=\linewidth]{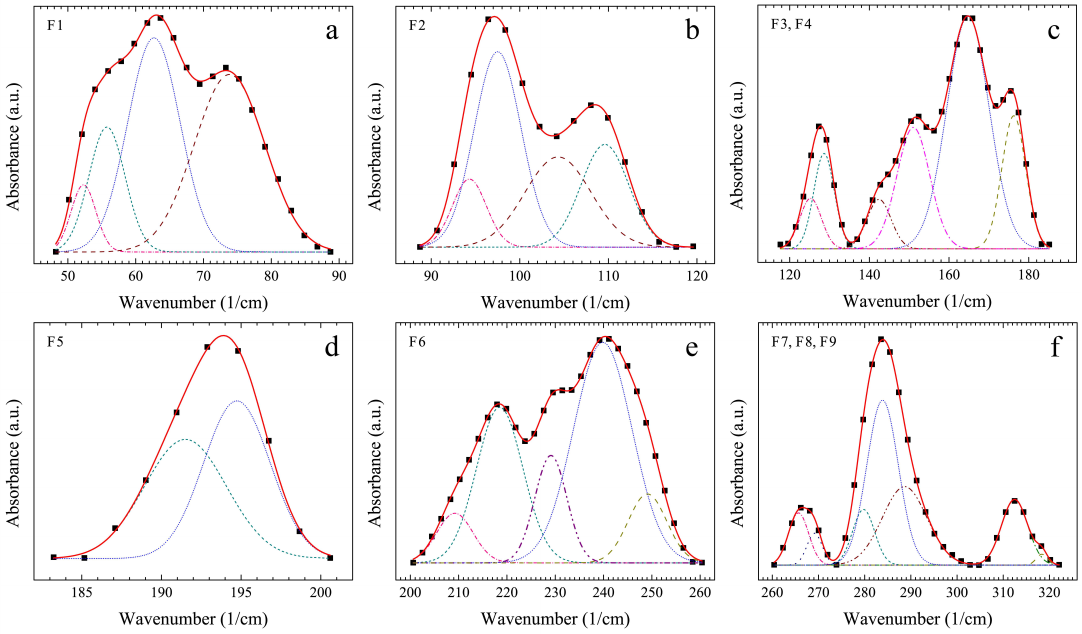}
	\caption{Absorption bands of orthorhombic KSbOSiO$_4$ in far-IR range and 
		their deconvolution using
		Voigt functions   
		\label{fig:FIR_bands}	 
	}
\end{figure}

Band F\,1 peaks at 63~cm$^{-1}$.
It comprises 2 local peaks at 63 ans 73~cm$^{-1}$, which consist of 4 lines with maximums at 52, 56, 63 and 74~cm$^{-1}$ (Fig.~\ref{fig:FIR_bands}\,a).
Band F\,2 peaking at 97~cm$^{-1}$ has 2 local peaks at 97 and 108~cm$^{-1}$; it comprises 4 components at 94, 97, 104 and 110~cm$^{-1}$ (Fig.~\ref{fig:FIR_bands}\,b).
The next band designated as F\,3 consists of a single maximum at 128~cm$^{-1}$; its peak analysis yields 2 constituting lines at 125 and 129~cm$^{-1}$;
deconvolution of band F\,4 with a maximum at 165~cm$^{-1}$ and three local peaks at 152, 165 and 175~cm$^{-1}$ reveals 4 components peaking at 142, 151, 165 and 176~cm$^{-1}$ (Fig.~\ref{fig:FIR_bands}\,c).
Band F\,5 with one maximum at 194~cm$^{-1}$ comprises 2 constituent lines at 192 and 195~cm$^{-1}$ (Fig.~\ref{fig:FIR_bands}\,d).
Band F\,6 reaching maximum at 240~cm$^{-1}$ with 2 local peaks at 218 and 240~cm$^{-1}$ comprises 5 components peaking at wavenumbers of 209, 219, 229, 240 and 249~cm$^{-1}$ (Fig.~\ref{fig:FIR_bands}\,e).
Finally, bands F\,7, F\,8 and F\,9 with maxima and local peaks at 267, 284 and 312~cm$^{-1}$, respectively, are composed by 2 lines at 266 and 269~cm$^{-1}$, 3 lines at 280, 284 and 289~cm$^{-1}$, and 2 lines at 312 and 319~cm$^{-1}$, respectively (Fig.~\ref{fig:FIR_bands}\,f).


\subsubsection{Raman spectrum and its constituent peaks}
\label{sec:Raman}

\begin{figure}[ht]
		\includegraphics*[width=\linewidth]{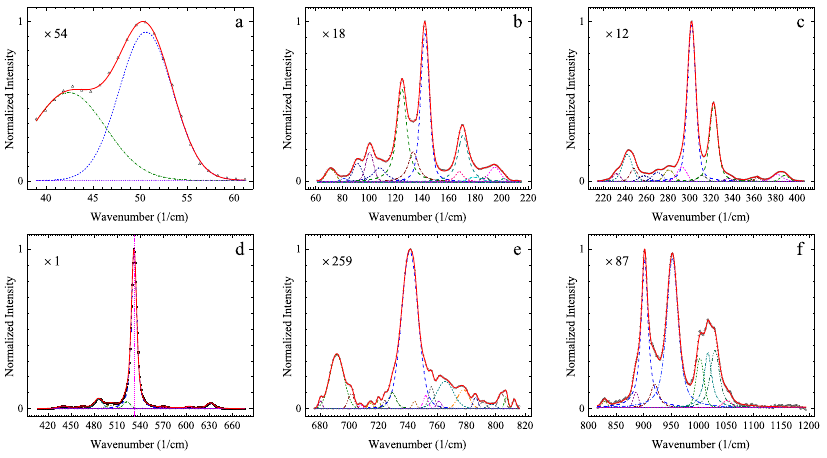}
	\caption{Raman bands of orthorhombic KSbOSiO$_4$ \cite{Beads_KSS_SPIE, KSS_Electron_microscopy} and 
		their deconvolution using Voigt functions; intervals from R\,1 (a) to R\,6 (f)
		\label{fig:Raman_bands}	 
	}
\end{figure}

We subdivided the Raman spectrum of orthorhombic KSS \cite{KSS_Raman_spectrum} into several intervals to analyze the peaks in each of them (Fig.~\ref{fig:Raman_bands}).
The list of lines derived from the spectrum by its deconvolution is tabulated in Table~\ref{tab:Raman-bands}.

The first interval ranges from 39 to 61~cm$^{-1}$ (Fig.~\ref{fig:Raman_bands}\,a);
there are 2 lines in this interval: one peaking at 42~cm$^{-1}$, and the other at 51~cm$^{-1}$.
The second band ranges from 61 to 214~cm$^{-1}$ (Fig.~\ref{fig:Raman_bands}\,b);
it comprises 8 maxima, the most intense of which is at a wavenumber of 142~cm$^{-1}$, and consists of 15 lines peaking at 
71, 81, 91, 100, 108, 125, 133, 142, 168, 171, 180, 187, 195, 202, and 212~cm$^{-1}$. 
The next interval extends from 214 to 406~cm$^{-1}$ (Fig.~\ref{fig:Raman_bands}\,c) and includes 7 maxima, the strongest of which is at a wavenumber of 302~cm$^{-1}$;
the peak analysis yields the following components of the spectrum segment peaking at the wavenumbers of 
223, 232, 242, 248, 258, 270, 281, 293, 302, 311, 322, 337, 347, 358, 363, 382, and 388~cm$^{-1}$.
Fig. ~\ref{fig:Raman_bands}\,d shows the region around the characteristic line in the range from 406 to 676~cm$^{-1}$;
the Raman lines detected in this interval of wavenumbers have maxima at 
440, 464, 486, 502, 523, 532, 543, 553, 570, 589, 604, 616, and 632~cm$^{-1}$.
The following segment ranging from 676 to 816~cm$^{-1}$ contains numerous weak components with the local maximum at a wavenumber of 742~cm$^{-1}$ (Fig.~\ref{fig:Raman_bands}\,e);
the lines have peaks at 
680, 692, 701, 708, 715, 720, 724, 729, 741, 745, 752, 756, 761, 765, 778, 786, 791, 797, 804, 807, and 813~cm$^{-1}$.
Finally, in the range from 816 to 1195~cm$^{-1}$, we detected two bands with comparable intensities peaking at 902 and 952~cm$^{-1}$ and a somewhat less intense one with a maximum at $\sim 1017.5$~cm$^{-1}$;
the lines that make up the spectrum peak at wavenumbers of 829, 886, 902, 920, 952, 1002, 1017, 1029, and 1052~cm$^{-1}$ in this range.%
\footnote{%
Two extra extremely weak lines were revealed at the wavenumber values of approximately 1282 and 1409~cm$^{-1}$ (\ref{ESM1}, Fig.~S1\,b), yet they are insufficiently trustworthy to be included in the list of KSS Raman lines.
}


\subsubsection{Brief discussion of vibrational spectra}
\label{sec:Disc}

The characteristic features of the vibrational spectra of KSS powder were previously discussed in Ref.~\cite{KSS_Raman&IR_spectra_Brandt}.
In this paper, in contrast to the previous study, the experimentally obtained spectra are carefully analyzed to identify the elementary vibrational peaks that comprise them by approximating the latter with Voigt functions.

The detailed deconvolution of the FTIR and Raman spectra of the synthesized orthorhombic KSS powder reveals a highly complex vibrational profile (Table~\ref{tab:spectr_compar}). 
The non-centrosymmetric orthorhombic lattice (\(Pna2_{1}\) space group) features a significantly distorted framework consisting of corner-sharing \([\text{SiO}_4]\) tetrahedra and \([\text{SbO}_6]\) octahedra. 
This structural topology, combined with the low-symmetry coordination spheres of the \(\text{K}^{+}\) cations, leads to a massive lifting of degeneracy, resulting in the high density of overlapping spectral components resolved in this work.


The M-range spectral envelope is dominated by the internal and bridging modes of the silicate anionic sublattices, highly coupled with the Sb--O framework vibrations.

The highest frequency bands, M\,8 (1085 to 1261~cm$^{-1}$) and M\,7 (879 to 1033~cm$^{-1}$), represent the asymmetric stretching vibrations of the silicate framework, namely \(\nu_{as}(\text{Si--O--Si})\) and \(\nu_{as}(\text{Si--O}^-)\) modes. 
Our peak analysis resolved 9 distinct sub-components in each of these bands. 
According to the density functional theory (DFT) calculations presented in Ref.~\cite{KSS_Raman&IR_spectra_Brandt}, this extensive splitting is a direct consequence of the in-equivalent Si--O bond lengths within the distorted $[\text{SiO}_4]$ tetrahedra, coupled with long-range electrostatic (LO--TO) splitting typical for polar orthorhombic KTP-type structures.

The intermediate frequency domain comprises bands M\,6 and M\,5 (691 to 773\,cm$^{-1}$). 
The solitary component at 773\,cm$^{-1}$ (M\,6) and the doublet at 691 and 710\,cm$^{-1}$ (M\,5) are assigned to the symmetric stretching vibrations of the bridging {Si--O--Si} bonds ($\nu_{s}(\text{SiOSi})$). 
Concurrently, this region gains intensity from the valence vibrations of the short, alternated antimonyl {Sb--O} bonds within the continuous {--Sb--O--Sb--O--} chains running parallel to the polar axes.

In the lower-energy mid-IR region, bands M\,2, M\,3, and M\,4 (421 to 634\,cm$^{-1}$) originate from a strong kinematic mixing between the internal bending modes of the \([\text{SiO}_4]\) tetrahedra ($\delta(\text{OSiO})$) and the stretching vibrations of the more rigid, heavier \([\text{SbO}_6]\) octahedra (\(\nu(\text{Sb--O})\)). 
The detection of 8 distinct constituent peaks within the M2 band profile underlines the deep structural distortion of the polyhedral network, which matches the predicted multi-component theoretical spectrum of Ref.~\cite{KSS_Raman&IR_spectra_Brandt}.

	
The F-range domain provides key insights into the low-frequency external lattice modes and inter-polyhedral motions.

Bands F\,7, F\,8, and F\,9 (266 to 319\,cm\(^{-1}\)) are predominantly governed by the collective bending deformations of the framework linkages, specifically the \(\delta(\text{O--Sb--O})\) and \(\delta(\text{O--Si--O})\) angles.
Moving lower in frequency, bands F4, F5, and F6 (142 to 249~cm\(^{-1}\)) are attributed to the translational and ``breathing'' vibrations of the heavier \([\text{SbO}_6]\) octahedral blocks acting as rigid units within the lattice.

The lowest energy acoustic-optical region, comprised of bands F\,1, F\,2, and F\,3 (52 to 129~cm\(^{-1}\)), samples the external lattice dynamics. 
Based on the DFT benchmarks \cite{KSS_Raman&IR_spectra_Brandt}, these multi-component peaks represent the external translatory modes of the heavy interstitial potassium (\(\text{K}^{+}\)) cations vibrating against the rigid anionic $(\text{SbOSiO}_4)^-$ framework. 
The presence of multiple lines (e.g., 4 components in F\,1 and 4 in F\,2) confirms that the \(\text{K}^{+}\) ions occupy highly asymmetric, multi-site crystalline positions within the channels of the orthorhombic lattice.


The deconvoluted Raman spectrum exhibits an exceptional match with the selection rules defined for the \(Pna2_{1}\) space group.

The spectral interval from 406 to 676~cm\(^{-1}\) contains the most crucial structural fingerprint of the material. 
As mentioned in Section~\ref{sec:samples}, the extraordinarily intense Raman line localized at 532\,cm\(^{-1}\) acts as a definitive spectroscopic marker for the orthorhombic phase of KSS. 
In accordance with Ref.~\cite{KSS_Raman&IR_spectra_Brandt}, this prominent mode is assigned to the totally symmetric valence vibration (A$_{1}$ or A$_1$-like) of the \([\text{SbO}_6]\) octahedron, where the shorter antimonyl bonds vibrate in-phase. 
The pristine profile, high relative intensity, and exact wavenumber alignment (532~cm\(^{-1}\)) of this line collectively demonstrate the macroscopic phase purity and high crystallinity of the KSS powder under study.

The weak, crowded Raman components resolved between 676 and 816\,cm$^{-1}$ (centered around 742\,cm$^{-1}$) mirror the high density of optical phonons predicted by theory. These correspond to the polar framework vibrations of the {Si--O--Si} bonds. 
Conversely, the high-frequency Raman domain (816 to 1195~cm\(^{-1}\)), containing components at 902, 952, and 1017.5~cm\(^{-1}\), originates from the symmetric and asymmetric valence stretching of the $[\text{SiO}_4]$ units. 
The notably suppressed intensity of these high-frequency Raman modes relative to their corresponding IR bands (M\,7, M\,8) perfectly aligns with the highly polar nature of the Si--O covalent bonds, which favor large dipole moment variations (IR active) over polarizability tensor changes (Raman active).

Finally, the extremely low-frequency Raman interval (39 to 214~cm\(^{-1}\)) successfully isolates the external acoustic-optical lattice modes. 
The resolved lines at 42 and 51~cm\(^{-1}\) and the multi-peak cluster up to 212~cm\(^{-1}\) represent the rotational (librational) modes of the polyhedral blocks intermixed with \(\text{K}^{+}\) translation vectors, providing a complete experimental verification of the low-frequency phonon density of states calculated for this system.


\subsection{Absorption and diffuse reflection spectra in the range from UV to near IR; band gap width}
\label{sec:Vis-UV}

\begin{figure}[t]
		\includegraphics*[width=\linewidth]{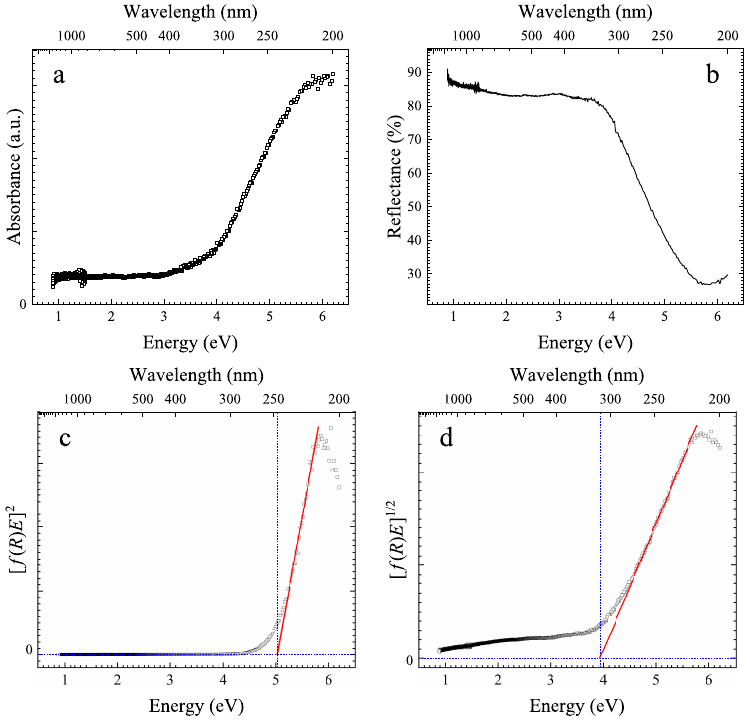}
	\caption{Absorption (a) and reflection (b) spectra of orthorhombic KSbOSiO$_4$ at $T=300$~K in the spectral range from 200 to 1400~nm 
		and 
		(c,\,d))
 estimation of the KSS band gap from the reflection spectrum obtained in the diffuse reflectance mode \cite{Optics_Light-scatter_materials, EB_Magnetic_Optical, Band-gap_energy_diffuse-1, Band-gap_energy_diffuse-2, EB_Photo@Cathodo}: $f(R) = (1 - R)^2/2R$, where $R$ is the reflection spectrum (b); the points of intersection of the straight lines fitting the linear parts of the functions 
 $[f(R)E]^2$ 
 (c)
 and
 $[f(R)E]^{1/2}$
 (d)
 with the $x$-axis ($E$) give 
 the allowed direct band gap width
 $E_{\mathrm{g}} = 5.02\pm 0.02$~eV 
 and
 the allowed indirect band gap width
 $E_{\mathrm{g_i}} = 3.93\pm 0.01$~eV
 (the confidence level is 95\,\% for both linear fits).		
		\label{fig:Vis-UV}	 
	}
\end{figure}

Absorption and reflection spectra of orthorhombic KSS powder, both recorded at a temperature of 300~K, are shown in Fig.~\ref{fig:Vis-UV}.
The absorption spectrum (Fig.~\ref{fig:Vis-UV}\,a) demonstrates a rapid growth of absorbance at photon energies $E\gtrsim 4$~eV (wavelengths $\lambda\lesssim 300$~nm).
Following Refs.~\cite{Optics_Light-scatter_materials, EB_Magnetic_Optical, Kubelka-Munk_function, Band-gap_energy_diffuse-1, Band-gap_energy_diffuse-2, Direct_indirect_band_UV-vis}, one can assess the values of allowed direct band gap ($E_{\mathrm{g}}$) 
and allowed indirect one ($E_{\mathrm{g_i}}$) 
from the diffuse reflection spectrum of KSS (Fig.~\ref{fig:Vis-UV}\,b) using the Kubelka-Munk model.

By plotting the graph of the function
 $[f(R)E]^2 = [(1 - R(E))^2E/2R(E)]^2$
(Fig.~\ref{fig:Vis-UV}\,c), where $R(E)$ is the reflection spectrum (Fig.~\ref{fig:Vis-UV}\,b), one can obtain the intersection point of the straight line approximating the linear part of the function 
$[f(R)E]^2$
and the abscissa axis $E$, which gives the value of the allowed direct band gap width $E_{\mathrm{g}}$. 
In the case of orthorhombic KSS, we have obtained   
$E_{\mathrm{g}} = 5.02\pm 0.02$~eV
(95\,\% confidence level).

Similarly, by plotting the graph of the function
$[f(R)E]^{1/2} = [(1 - R(E))^2E/2R(E)]^{1/2}$
(Fig.~\ref{fig:Vis-UV}\,d), one can obtain the intersection point of the straight line approximating the linear part of the function 
$[f(R)E]^{1/2}$
and the abscissa axis $E$, which gives the value of the width of the allowed indirect band gap 
$E_{\mathrm{g_i}} = 3.93\pm 0.01$~eV
(95\,\% confidence level).%

It should be emphasized that
extrapolation of the linear segment of the $[f(R)E]^{1/2}$ Tauc plot (Fig.~\ref{fig:Vis-UV}\,d) to the photon energy axis ($(Y=0)$) yields a {\it fundamental} indirect band gap of $E_{\mathrm{g_i}} = 3.93\pm 0.01$~eV. 
The gradual, slight ascent observed below this threshold (in the range from 1 to 4~eV ) is attributed to sub-bandgap absorption within the defect-induced states (Urbach tail) and light scattering effects. 
It is worth noting that alternative approaches, such as the two-line intersection method frequently employed for highly disordered systems like metal-organic frameworks (MOFs) \cite{Direct_indirect_band_UV-vis}, shift the apparent absorption edge toward 4.1--4.2~eV in this case. 
For a crystalline oxide matrix like KSbOSiO$_{4}$, however, such an approach artificially overestimates the true threshold of interband transitions, whereas the classical linear extrapolation to zero absorption remains the most physically grounded method.

The observed substantial energy separation (approximately 1.1~eV) between the indirect ($E_{\mathrm{g_i}} = 3.93$~eV) and direct ($E_{\mathrm{g}} = 5.02$~eV) band gaps is intrinsically linked to the crystallographic and orbital architecture of the orthorhombic KSS lattice. 
In this framework structure, the top of the valence band is predominantly governed by the non-bonding O~\(2p\) states (lone pairs), meaning that the highest occupied electronic density remains strongly localized on the oxygen anions.
Within the crystal framework, these oxygen atoms serve as bridging ligands, forming both rigid, highly covalent \([\text{SiO}_{4}]\) tetrahedra and more ionic \([\text{SbO}_{6}]\) octahedra \cite{KSS_Raman&IR_spectra_Brandt}. 
This heterogeneous coordination environment and asymmetric chemical bonding cause the electronic wave functions at the valence band maximum to exhibit pronounced energy dispersion depending on the crystallographic direction within the reciprocal \(k\)-space. 
Consequently, the valence band maximum shifts away from the center of the Brillouin zone (the \(\Gamma\) point).

Conversely, the bottom of the conduction band is predominantly formed by the highly dispersed states originating from the Sb~\(5d\) (or partially Sb~\(5p\)) orbitals within the \([\text{SbO}_{6}]\) octahedra, with a higher-energy contribution from the unoccupied states of the \(sp^{3}\)-hybridized Si atoms within the \([\text{SiO}_{4}]\) tetrahedra.
Due to the spatial alternation of these polyhedra, the orbital overlap trajectories vary drastically along different crystallographic directions (e.g., $[001]$, $[100]$). 
As a result, the conduction band minimum emerges at a completely distinct point in the $k$-space compared to the valence band maximum.

Thus, the pronounced structural anisotropy and complex orbital overlapping inherent to the orthorhombic symmetry force the valence band maximum and the conduction band minimum to settle at entirely different $k$-points within the Brillouin zone. 
Consequently, the lowest-energy fundamental transition requires phonon participation to compensate for the momentum mismatch, while the direct vertical electronic transition demands a significantly higher photon energy of 5.02~eV.

\subsection{CL and PL spectra of orthorhombic KSS}
\label{sec:lumin}

\subsubsection{CL spectra at 300 and 80~K and their peak analysis}
\label{sec:CL}

\begin{figure}[ht]
	\begin{minipage}[l]{1\textwidth}{\vspace{-0cm}}
		\begin{tabular}{l}
			\includegraphics[width=0.7\textwidth]{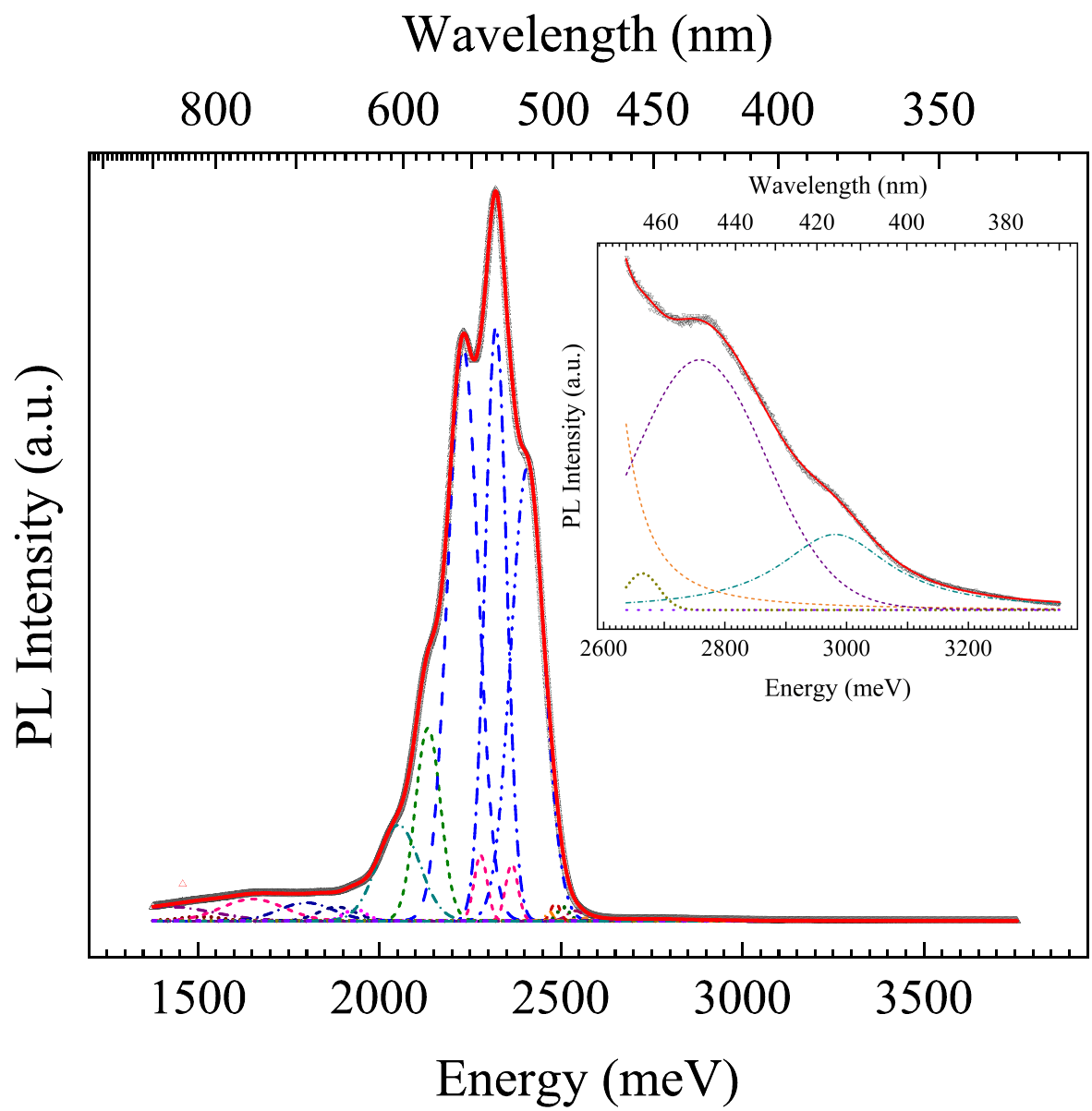}
		\end{tabular}
	\end{minipage}
	\caption{PL band of orthorhombic KSbOSiO$_4$ and 
		its deconvolution using Voigt functions: $T=300$~K;
		$\lambda_{\mathrm{ex}}=325$~nm ($E_{\rm b}\approx3.81$~eV);
		the insert shows the high-energy wing of the band. 
		\label{fig:PL_band} 
	}
\end{figure}

CL spectra of orthorhombic KSS powder obtained at 300 and 80~K were presented, likely for the first time, in Ref.~\cite{KSS_Electron_microscopy}.
They were used when studying the effect of KSS microcrystals in historical turquoise glass beads on their corrosion induces by the internal stress, and no data on constituent peaks of these bans were presented in that work. 
In this Section, we analyze these peaks for the band composition (Fig.~\ref{fig:PL_band}). 
The main characteristics of the revealed peaks are listed in Table~\ref{tab:CL-bands}.

The CL bands exhibit two distinct maxima at 2.79 and 2.57~eV (444 and 483~nm) both at 80 and 300~K.
These CL bands can be used as characteristic bands of orthorhombic KSS for the spectral analysis.
Peak analysis has revealed 8 peaks constituting the bands recorded at room temperature and at liquid nitrogen temperature; they are as follows (Table~\ref{tab:CL-bands}):
the peak at $\sim2.55$~eV is the strongest one,
those at 2.78 and $\sim2.93$~eV are somewhat less intense ones,
the next in intensity one peaks at 2.34~eV,
and
the weakest peaks have maxima at
$\sim1.75$
$\sim2.06$,
$\sim2.20$
and
$\sim3.20$~eV.

It should be recalled that the above CL peaks were excited by electrons with the energy of 15~keV, 
which throw electrons of the crystal from the valence band high into the conduction band, from where they non-radiatively transit to the bottom of the conduction band, emitting a great number of phonons, via numerous transitions in the valence band, and only then recombine through localized states in the band gap 
($E_{\mathrm{g}} \approx 5.02$~eV, $E_{\mathrm{g_i}} \approx 3.93$~eV), 
related to the levels of impurities and defects of the KSS crystal lattice, with the emission of photons.

Since we were specifically studying powder, heat transfer from the recording location, even at a microscope stage temperature of 80~K, could have been quite weak. 
Consequently, the spectra recording location could have heated up at both the nominal temperature of 300 and 80~K.
The real temperature at this location could have been significantly higher than the nominal one in both cases.
Nevertheless, it should be noted that KSS microcrystals embedded in glass exhibit virtually the same CL spectra as those registered from powder, both at the room temperature and the liquid nitrogen temperature \cite{KSS_Electron_microscopy}, although the heat transfer in glass is undoubtedly much higher than in powder under high vacuum conditions.
Thus, we can suggest that the revealed broad band likely associated with deep levels of defects and impurities (see, e.g., \cite{2024-Tomography_CdZnSSe_beads, LSM-2018, ZnSe:Fe_spatial-2019, PL_ZnO_pigment_Zn-complexes, PL_ZnO_films_different_deposition, CaCuSi4O10_Falk}) is virtually independent of the crystal temperature, with the possible exception of its intensity \cite{KSS_PL}. 

It should also be noted that our data are in fairly good agreement with old measurements of KSS PL under ultraviolet excitation carried out at a temperature of 4.2~K \cite{KSS_PL}.  
The emission spectrum of KSS, which consisted of a broad band with a maximum at about 470~nm, was presented in that work, while the excitation spectrum consisted of a broad band peaked at a wavelength of about 240~nm ($E \approx 5.16$~eV that exceeds the value of 
$E_{\mathrm{g}}\approx 5.02$~eV 
obtained in this work at room temperature\,---\,see Section~\ref{sec:Vis-UV}).

Assuming that the excitation peak registered in Ref.~\cite{KSS_PL} at \(T\approx4.2\)~K is associated with the band gap, i.e.  
$E_{\mathrm{g}}(0\,K)\approx E_{\mathrm{g}}(4.2\,K)\sim 5.16$~eV,
the mean value of $dE_{\rm g}/dT$ can be roughly estimated from the above data of this work and Ref.~\cite{KSS_PL} as 
$\sim-4.7\times 10^{-4}$~eV/K.

\subsubsection{PL spectrum at room temperature and its peak analysis}
\label{sec:PL}

Since we recorded the PL spectra under laser excitation with a wavelength $\lambda_{\mathrm{ex}}=325$~nm (that corresponded to the photon energy $E_{\rm ex}\approx3.81$~eV, which was noticeably lower than $E_{\rm g}$ but close to $E_{\rm g_i}$), we obtained a PL spectrum consisting of the emission due only to transitions from levels located in the band gap, the energy of which was lower than the energy of the exciting beam (Fig.~\ref{fig:PL_band}). 
Nonetheless, peak analysis of the obtained band with two distinct local maxima yielded a reach set of intense peaks accompanied with numerous weak and very weak lines (Table~\ref{tab:PL-bands}). 

The three strongest peaks are observed at energies of 
2.23, 2.32 and, somewhat less intense, 2.41~eV.
Two weaker peaks are visible at 2.05 and 2.13~eV.
The rest weak lines are presented in Table~\ref{tab:PL-bands}.

The strongest CL peak at $\sim2.55$~eV corresponds to the weak PL line at 2.54~eV.
The intense CL peaks at 2.78 and $\sim2.93$~eV are seen as the weak line at 2.76~eV and the very weak line at $\sim2.98$~eV in the PL spectrum.
The CL peak at 2.34~eV matches to the PL one at 2.32~eV.
Finally, the CL peaks at 2.06 and 1.75 eV may possibly be associated with the PL peaks at 2.05 and 1.79~eV.
The CL peak at $\sim2.20$~eV is most likely masked by the intense PL peak located in the same spectral region.
The peak at $\sim3.20$~eV, present in the CL spectrum, is absent in the PL spectrum.

\section{Conclusion }
\label{sec:concl}

In conclusion of the article, we consider it necessary to highlight the main provisions of this work.

Firstly, a detailed analysis of the IR absorption and Raman spectra of a specially synthesized crystalline powder of orthorhombic KSS allowed us to perform a comprehensive spectroscopic assignment of deconvoluted IR and Raman bands in the spectral range from 50 to 4000~cm$^{-1}$ and compare them with literature data.

Secondly, the analysis of diffuse reflection spectrum in the range from near infrared to ultraviolet have enabled the assessment of the widths of allowed direct and allowed indirect band gaps of orthorhombic KSS using the Kubelka-Munk model.
We have found the direct band gap to be
$E_{\mathrm{g}} = 5.02\pm 0.02$~eV
and the indirect one to be 
$E_{\mathrm{g_i}} = 3.93\pm 0.01$~eV
(confidence level is 95\,\% for both estimates).

Thirdly, we can conclude that the cathodoluminescence and photoluminescence spectra studied and analyzed in this work should be assigned to the light emission caused by charge carrier transitions from deep levels of intrinsic defects and/or impurities.

Finally, we succeeded to roughly estimate the mean value of the temperature coefficient of the direct band gap width in the interval from 0 to 300 K as $dE_{\rm g}/dT\sim-4.7\times 10^{-4}$~eV/K.

 \section*{CRediT authorship contribution statement}
 
 \textbf{Vladimir A. Yuryev:}
 Conceptualization,
 Methodology,
 Funding acquisition, 
 Investigation,
 Data Curation,
 Formal analysis,
 Writing\,--
 Original draft, 
 Writing\,-- Review
 \& Editing,
 Supervision.
 \textbf{Tatyana V. Yuryeva:}
 Conceptualization,
 Funding acquisition,
 Project administration,
 Resources,
 Data Curation,
 Investigation,
 Visualization.
 \textbf{Ekaterina A. Morozova:}
 Investigation.
 \textbf{Irina F. Kadikova:}
 Investigation.
 \textbf{Polina A. Edelbekova:}
 Investigation.

 
 \section*{Declaration of competing interest} 
 
 The authors declare that they have no known competing financial interests or personal relationships that could have appeared to influence the work reported in this paper.
 
 \section*{Data availability} 
 The data required to reproduce the findings of this study are available from the corresponding author upon request.




\section*{Acknowledgments}


We thank 
Mr. Ilya~B.~Afanasyev from
Professor A.\,R.\,Shlyakhov Russian Federal Center of Forensic Science of the Ministry of Justice 
for the SEM images and
Dr. Sergey~A.~Malyhin from 
P.~N.~Lebedev Physical Institute of the Russian Academy of Sciences
for the CL measurements.
%

\clearpage



\clearpage

\section*{Figures and Tables}

\begin{table}[th] {\vspace{-2.5cm}} 
	\caption{Spectral bands of mid-IR radiation absorption in orthorhombic KSbOSiO$_4$
		\label{tab:Mid-IR_bands}	 
	}
	\begin{minipage}{\textwidth} 
		\begin{tabular}{lllllll}
			\noalign{\smallskip}\hline
			\multicolumn{3}{l}{Band}& Local peaks &\multicolumn{3}{l}{Deconvolution\footnote{\label{foot-tab1:Voigt} Voigt peak profiles are implied.}} \\
			\cline{1-3} \cline{5-7}
			Number & Maximum & Intensity &    & Peak centers & Peak heights & FWHM \\ \hline
			& (cm$^{-1}$) & (rel. units) &  (cm$^{-1}$) &  (cm$^{-1}$)& (rel. units)&  (cm$^{-1}$)\\
			\hline
			M\,1& 358 & 1.00  &  358; 378   & 338; 358;   & 0.07; 0.99; & 13.7; 22.5;\\
			 	 &     &       &             &  381        &  0.51 		& 16.5\\
			M\,2& 466 & 1.00  & 466; 485;   & 421; 446;   & 0.09; 0.53; & 22.1; 32.2;\\
				 &     &       &  504        & 451; 469;   & 0.04; 0.86; & 9.3; 26.9; \\
				 &     &       &             & 488; 505;   & 0.33; 0.59; & 11.9; 19.3;\\
				 &     &       &             & 521; 533    & 0.14; 0.04  & 15.1; 16.8 \\
			M\,3& 593 & 0.51  & 576; 593    & 550; 569;   & 0.01; 0.13; & 3.1; 13.8; \\
			&     &      &                  & 575; 592;	 & 0.15; 0.48  & 8.0; 25.2; \\
			&     &      &                  & 603  	    & 0.02        & 6.2   		  \\
			M\,4& 624 & 0.68  & 624         & 621; 628;    & 0.36; 0.07;& 9.8; 6.7;\\
			&     &      &                  & 634  	    & 0.53        & 25.6   		  \\
			M\,5& 710 & 0.56  & 710         & 691; 710    & 0.04; 0.54  & 16.9; 22.7 \\ 
		%
			M\,6& 773 & 0.29  & 773         & 773         & 0.29 		   & 19.1\\
			M\,7& 915 & 0.81  & 897; 915;   & 879; 886;   & 0.43; 0.15; & 71.6; 27.7;\\
			&&&					  950         & 897; 916;   & 0.13; 0.27; & 14.8; 22.6;\\
			&&&							    	  & 933; 945;   & 0.05; 0.50; & 12.3; 74.7;\\			
			&&&							    	  & 954; 1006;  & 0.05; 0.07; & 12.9; 44.9;\\
			&&&							    	  & 1033        & 0.01 			& 13.0\\
			M\,8& 1093 & 0.031 & 1093; 1194; & 1085; 1105; & 0.018; 0.014;   & 24.0; 27.6; \\
			&&                & 1203; 1261  & 1133; 1200; & 0.018; 0.016;   & 76.0; 57.4; \\
			&&                &     		  & 1203; 1209; & 0.0025; 0.0019; & 4.1; 7.2;   \\
			&&                &     		  & 1221; 1237; & 0.0014; 0.0028; & 7.4; 27.5;  \\			
			&&                &     		  & 1261        & 0.0036          & 9.0 \\
			\hline
		\end{tabular}
	\end{minipage}
\end{table}


\begin{table}[th]
	\caption{Spectral bands of far-IR radiation absorption in orthorhombic KSbOSiO$_4$
		\label{tab:Far-IR_bands}	 
	}
	\begin{minipage}{\textwidth}
		\begin{tabular}{lllllll}
			\noalign{\smallskip}\hline
		\multicolumn{3}{l}{Band}& Local peaks &\multicolumn{3}{l}{Deconvolution\footnote{\label{foot-tab2:Voigt} Voigt peak profiles are implied.}} \\
		\cline{1-3} \cline{5-7}
		Number & Maximum & Intensity &    & Peak centers & Peak heights & FWHM \\ \hline
		& (cm$^{-1}$) & (rel. units) &  (cm$^{-1}$) &  (cm$^{-1}$)& (rel. units)&  (cm$^{-1}$)\\
		\hline
			F\,1& 63  & 1.00 & 63; 73    & 52; 56;   & 0.29; 0.55; & 4.5; 6.4; \\
			&&&           					  & 63; 74    & 0.92; 0.78  & 9.0; 12.7 \\
			F\,2& 97  & 0.70 & 97; 108   & 94; 97;   & 0.20; 0.59; & 4.6; 6.3; \\
			&&&                          & 104; 110  & 0.27; 0.31  & 9.0; 6.5  \\
			F\,3& 128 & 0.25 & 128       & 125; 129  & 0.10; 0.19  & 6.4; 6.1 \\
			F\,4& 165 & 0.46 & 152; 165; & 142; 151; & 0.10; 0.24; & 7.0; 9.7; \\
			&&	 				  & 175 		  & 165; 176  & 0.47; 0.27  & 12.5; 7.1 \\	
			F\,5& 194 & 0.25 & 194       & 192; 195  & 0.14; 0.18  & 6.4; 5.1 \\
			F\,6& 240 & 0.67 & 218; 240  & 209; 219; & 0.16; 0.45; & 9.8; 11.4; \\
			&     &      &               & 229; 240; & 0.32; 0.64; & 8.0; 14.4; \\
			&     &      &               & 249       & 0.21  		 & 10.2       \\
			F\,7& 267 & 0.15 & 267       & 266; 269; & 0.14; 0.09; & 5.4; 4.1  \\
			F\,8& 284 & 0.61 & 284       & 280; 284; & 0.15; 0.44 & 5.7; 8.2; \\
			&     &      &               & 289       & 0.21       & 11.8      \\
			F\,9& 312 & 0.18 & 312       & 312; 319  & 0.18; 0.03 & 6.8; 3.0  \\
			%
			\hline
		\end{tabular}
	\end{minipage}
\end{table}


\begin{table}[th]{\vspace{-3.5cm}}
	\caption{%
		Properties of the Raman peaks of orthorhombic KSS derived from the spectral bands plotted in Fig.~\ref{fig:Raman_bands}. 
		\label{tab:Raman-bands}
	}
		{\hspace{-2cm}}
		\begin{tabular}{ccccccccccc}
			\noalign{\smallskip}\hline 
			Center  &Intensity	&FWHM 	 &&	Center  	&Intensity		&FWHM  &&	Center  	&Intensity		&FWHM\\ 								
			(cm$^{-1}$)	  &(rel. units)&(cm$^{-1}$)	 	 &&(cm$^{-1}$)	  		&(rel. units)	&(cm$^{-1}$)	&&(cm$^{-1}$)	  		&(rel. units)	&(cm$^{-1}$) \\ 
			\noalign{\smallskip}\hline \noalign{\smallskip}
			$42 $&$ 1.03\times 10^{-2} $&$ 9.2 $&$  $&$ 311 $&$ 3.03\times 10^{-3} $&$ 9.4 $&$  $&$ 720 $&$ 2.11\times 10^{-4} $&$ 2.8 $\\
			$51 $&$ 1.74\times 10^{-2} $&$ 6.7 $&$  $&$ 322 $&$ 3.90\times 10^{-2} $&$ 9.0 $&$  $&$ 724 $&$ 2.03\times 10^{-4} $&$ 3.2 $\\
			$71 $&$ 4.39\times 10^{-3} $&$ 8.5 $&$  $&$ 337 $&$ 1.47\times 10^{-3} $&$ 9.3 $&$  $&$ 729 $&$ 3.96\times 10^{-4} $&$ 6.1 $\\
			$81 $&$ 1.14\times 10^{-3} $&$ 7.4 $&$  $&$ 347 $&$ 9.51\times 10^{-5} $&$ 2.4 $&$  $&$ 741 $&$ 3.85\times 10^{-3} $&$ 13.2 $\\
			$91 $&$ 6.47\times 10^{-3} $&$ 8.1 $&$  $&$ 358 $&$ 8.50\times 10^{-4} $&$ 6.9 $&$  $&$ 745 $&$ 1.91\times 10^{-4} $&$ 3.3 $\\
			$100 $&$ 1.02\times 10^{-2} $&$ 7.1 $&$  $&$ 363 $&$ 1.84\times 10^{-3} $&$ 6.5 $&$  $&$ 752 $&$ 3.23\times 10^{-4} $&$ 4.0 $\\
			$108 $&$ 4.92\times 10^{-3} $&$ 11.8 $&$  $&$ 382 $&$ 3.51\times 10^{-3} $&$ 11.5 $&$  $&$ 756 $&$ 5.39\times 10^{-4} $&$ 4.3 $\\
			$125 $&$ 3.22\times 10^{-2} $&$ 8.9 $&$  $&$ 388 $&$ 3.01\times 10^{-3} $&$ 9.7 $&$  $&$ 761 $&$ 1.88\times 10^{-4} $&$ 4.0 $\\
			$133 $&$ 1.05\times 10^{-2} $&$ 8.8 $&$  $&$ 440 $&$ 1.52\times 10^{-2} $&$ 17.1 $&$  $&$ 765 $&$ 6.73\times 10^{-4} $&$ 13.3 $\\
			$142 $&$ 5.15\times 10^{-2} $&$ 7.8 $&$  $&$ 464 $&$ 1.30\times 10^{-2} $&$ 13.4 $&$  $&$ 778 $&$ 4.66\times 10^{-4} $&$ 9.2 $\\
			$168 $&$ 3.63\times 10^{-3} $&$ 7.1 $&$  $&$ 486 $&$ 5.85\times 10^{-2} $&$ 12.5 $&$  $&$ 786 $&$ 2.82\times 10^{-4} $&$ 3.9 $\\
			$171 $&$ 1.63\times 10^{-2} $&$ 9.3 $&$  $&$ 502 $&$ 2.34\times 10^{-2} $&$ 16.5 $&$  $&$ 791 $&$ 1.62\times 10^{-4} $&$ 5.1 $\\
			$180 $&$ 1.70\times 10^{-3} $&$ 12.1 $&$  $&$ 523 $&$ 4.46\times 10^{-2} $&$ 11.0 $&$  $&$ 797 $&$ 1.11\times 10^{-4} $&$ 6.9 $\\
			$187 $&$ 1.82\times 10^{-3} $&$ 6.6 $&$  $&$ 532 $&$ 1.00 \times 10^{0~\,}    $&$ 8.4 $&$  $&$ 804 $&$ 3.72\times 10^{-4} $&$ 5.1 $\\
			$195 $&$ 5.12\times 10^{-3} $&$ 10.0 $&$  $&$ 543 $&$ 5.51\times 10^{-3} $&$ 7.2 $&$  $&$ 807 $&$ 3.01\times 10^{-4} $&$ 2.8 $\\
			$202 $&$ 1.14\times 10^{-3} $&$ 5.8 $&$  $&$ 553 $&$ 3.16\times 10^{-3} $&$ 12.2 $&$  $&$ 813 $&$ 2.34\times 10^{-4} $&$ 2.8 $\\
			$212 $&$ 1.12\times 10^{-4} $&$ 2.7 $&$  $&$ 570 $&$ 1.07\times 10^{-2} $&$ 13.9 $&$  $&$ 829 $&$ 4.55\times 10^{-4} $&$ 11.2 $\\
			$223 $&$ 4.03\times 10^{-4} $&$ 5.4 $&$  $&$ 589 $&$ 3.41\times 10^{-3} $&$ 11.9 $&$  $&$ 886 $&$ 1.12\times 10^{-3} $&$ 11.3 $\\
			$232 $&$ 3.90\times 10^{-3} $&$ 8.4 $&$  $&$ 604 $&$ 1.33\times 10^{-2} $&$ 10.5 $&$  $&$ 902 $&$ 1.08\times 10^{-2} $&$ 14.4 $\\
			$242 $&$ 1.39\times 10^{-2} $&$ 10.8 $&$  $&$ 616 $&$ 3.75\times 10^{-3} $&$ 9.4 $&$  $&$ 920 $&$ 1.71\times 10^{-3} $&$ 17.7 $\\
			$248 $&$ 6.55\times 10^{-3} $&$ 8.1 $&$  $&$ 632 $&$ 3.75\times 10^{-2} $&$ 12.4 $&$  $&$ 952 $&$ 1.08\times 10^{-2} $&$ 24.9 $\\
			$258 $&$ 2.90\times 10^{-3} $&$ 11.5 $&$  $&$ 680 $&$ 1.95\times 10^{-4} $&$ 2.8 $&$  $&$ 1002 $&$ 3.57\times 10^{-3} $&$ 16.7 $\\
			$270 $&$ 4.72\times 10^{-3} $&$ 11.1 $&$  $&$ 692 $&$ 1.33\times 10^{-3} $&$ 11.4 $&$  $&$ 1017 $&$ 4.01\times 10^{-3} $&$ 16.2 $\\
			$281 $&$ 5.99\times 10^{-3} $&$ 10.3 $&$  $&$ 701 $&$ 3.72\times 10^{-4} $&$ 4.9 $&$  $&$ 1029 $&$ 4.16\times 10^{-3} $&$ 20.1 $\\
			$293 $&$ 7.45\times 10^{-3} $&$ 10.3 $&$  $&$ 708 $&$ 9.75\times 10^{-5} $&$ 2.9 $&$  $&$ 1052 $&$ 5.35\times 10^{-4} $&$ 18.9 $\\
			$302 $&$ 7.98\times 10^{-2} $&$ 9.2 $&$  $&$ 715 $&$ 1.48\times 10^{-4} $&$ 3.9 $&$  $&$  $&$  $&$  $\\
			\noalign{\smallskip}
			\hline	
		\end{tabular} \normalsize  
\end{table}

\begin{table*}[th]{\vspace{-3.5cm}}
	\caption{Comprehensive spectroscopic assignment and comparison of deconvoluted IR and Raman bands for orthorhombic KSS against literature benchmarks \cite{KSS_Raman&IR_spectra_Brandt}.}
	\label{tab:spectr_compar}
{\hspace{-4cm}}
	\footnotesize
	\begin{tabular}{cccccc}
		\noalign{\smallskip}\hline 
		Spectral Range & Band ID & Experimental Maxima & Resolved Components  & {Ref.~\cite{KSS_Raman&IR_spectra_Brandt}} & Primary Vibrational Assignment \\
		&&(cm$^{-1}$) &(cm$^{-1}$)&(cm$^{-1}$)&\\
		\noalign{\smallskip}\hline \noalign{\smallskip}
		{Mid-IR (M)} & M\,8 & 1093 & 1085, 1105, 1133, 1200, 1203, 1209, 1221, 1237, 1261 & 1080--1260 & $\nu_{as}(\text{Si--O--Si})$ and $\nu_{as}(\text{Si--O}^-)$ \\
		(325--1300 cm$^{-1}$) & M\,7 & 915  & 879, 886, 897, 916, 933, 945, 954, 1006, 1033 & 880--1040  & $\nu_{as}(\text{Si--O})$ stretching \\
		& M\,6 & 773  & 773 & 775 & $\nu_{s}(\text{Si--O--Si})$ / short $\text{Sb--O}$ \\
		& M\,5 & 710  & 691, 710 & 695, 712 & $\nu_{s}(\text{Si--O--Si})$ stretching \\
		& M\,4 & 624  & 621, 628, 634 & 620--635 & $\delta(\text{O--Si--O})$ / $\nu(\text{Sb--O})$ mix \\
		& M\,3 & 593  & 550, 569, 575, 592, 603 & 550--605 & $\delta(\text{O--Si--O})$ deformation \\
		& M\,2 & 466  & 421, 446, 451, 469, 488, 505, 521, 533 & 420--530 & Internal $[\text{SiO}_4]$ bending \\
		& M\,1 & 358  & 338, 358, 381 & 340--380 & $\nu(\text{Sb--O})$ octahedral stretching \\
		{Far-IR (F)} & F\,9 & 312  & 312, 319 & 310, 320 & $\delta(\text{O--Sb--O})$ / $\delta(\text{O--Si--O})$ \\
		(40--325 cm$^{-1}$)  & F\,8 & 284  & 280, 284, 289 & 280--290 & Framework angular deformations \\
		& F\,7 & 267  & 266, 269 & 265--270 & Inter-polyhedral bending \\
		& F\,6 & 240  & 209, 219, 229, 240, 249 & 210--250 & $[\text{SbO}_6]$ octahedral breathing \\
		& F\,5 & 194  & 192, 195 & 193, 196 & Rigid octahedral translations \\
		& F\,4 & 165  & 142, 151, 165, 176 & 140--175 & Translational matrix modes \\
		& F\,3 & 128  & 125, 129 & 126, 130 & $\text{K}^+$ cation translations \\
		& F\,2 & 97   & 94, 97, 104, 110 & 95--112 & External $\text{K}^+$ lattice modes \\
		& F\,1 & 63   & 52, 56, 63, 74 & 50--75 & Low-frequency acoustic-optical \\
		{Raman}      & R\,6 & --   & 829, 886, 902, 920, 952, 1002, 1017, 1029, 1052 & 830--1060 & Valence stretching of $[\text{SiO}_4]$ \\
		(39--1195 cm$^{-1}$) & R\,5 & 742  & 680, 692, 701, 708--807, 813 (21 lines) & 680--815 & $\text{Si--O--Si}$ bond modes \\
		& R\,4 & \textbf{532} & 440, 464, 486, 502, 523, \textbf{532}, 543--632 & \textbf{532} (strong) & $\text{A}_1$ symmetric $\nu(\text{Sb--O})$ \textit{(Marker)} \\
		& R\,3 & 302  & 223, 232, 242, 248, 258, 270--388 (17 lines) & 220--390 & Mixed deformations / translations \\
		& R\,2 & 142  & 71, 81, 91, 100, 108, 125, 133, 142--212 & 70--215 & Polyhedral rotations / $\text{K}^+$ translation \\
		& R\,1 & --   & 42, 51 & 40, 50 & Acoustic-optical lattice vibrations \\
					\noalign{\smallskip}
		\hline	
	\end{tabular} 
\end{table*}


\begin{figure}[ht]
		\includegraphics*[width=\linewidth]{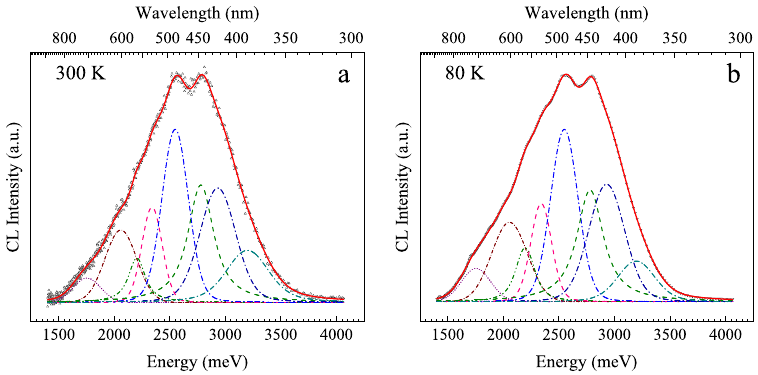}
	\caption{CL band of orthorhombic KSbOSiO$_4$ \cite{KSS_Electron_microscopy} and 
		its deconvolution using Voigt functions: $T=300$~(a) and 80~K (b);
		$E_{\rm b}=15$~keV
		\label{fig:CL_bands} 
	}
\end{figure}

\begin{table}[th]{\vspace{-0cm}}
	\caption{%
		Properties of the CL peaks of orthorhombic KSS derived from the spectra plotted in Fig.~\ref{fig:CL_bands};
		the intensity is normalized relative to the maximum intensity of the	cumulative band. 
		\label{tab:CL-bands}
	}
		\begin{tabular}{ccccccccc}
			\noalign{\smallskip}\hline 
			%
			Center  &Intensity	&FWHM 	 & Area		&&	Center  &Intensity	&FWHM  & Area\\ 								
			(eV)	  &(rel. units)&(eV)	 	 & (\%) 		&&(eV)	  &(rel. units)&(eV)	 & (\%) 	\\ 
			\noalign{\smallskip} \hline \noalign{\smallskip}
			\multicolumn{4}{c}{300~K, panel\,(a)}	&& \multicolumn{4}{c}{80~K, panel\,(b)} \\
			\noalign{\smallskip}\hline \noalign{\smallskip}
			1.755 & 0.109 & 0.291 & 3.07 				&& 1.754 & 0.149 & 0.312 & 4.35  \\
			2.059 & 0.323 & 0.337 & 10.59				&& 2.055 & 0.355 & 0.380 & 12.76  \\
			2.204 & 0.198 & 0.180 & 4.71 				&& 2.196 & 0.241 & 0.202 & 5.73  \\
			2.340 & 0.425 & 0.216 & 9.20 				&& 2.340 & 0.439 & 0.222 & 9.65  \\
			2.548 & 0.773 & 0.280 & 21.90 			&& 2.548 & 0.770 & 0.291 & 21.94  \\
			2.780 & 0.524 & 0.279 & 19.06 			&& 2.780 & 0.499 & 0.287 & 18.29  \\
			2.932 & 0.511 & 0.382 & 20.06 			&& 2.929 & 0.525 & 0.384 & 19.92  \\
			3.199 & 0.232 & 0.453 & 11.41 			&& 3.196 & 0.182 & 0.397 & 7.37  \\
			\hline	
		\end{tabular} 
\end{table}


\begin{table}[th]{\vspace{-0cm}}
	\caption{%
		Properties of the PL peaks of orthorhombic KSS derived from the spectrum plotted in Fig.~\ref{fig:PL_band};
		the intensity is normalized relative to the maximum intensity of the	cumulative band. 
		\label{tab:PL-bands}
	}
		\begin{tabular}{ccccccccc}
			\noalign{\smallskip}\hline 
			%
			Center  &Intensity	&FWHM 	 & Area		&&	Center  &Intensity	&FWHM  & Area\\ 								
			(eV)	  &(rel. units)&(eV)	 	 & (\%) 		&&(eV)	  &(rel. units)&(eV)	 & (\%) 	\\ 
			\noalign{\smallskip} \hline \noalign{\smallskip}
			1.414 & 0.0196 & 0.297 & 1.335 				&& 2.365 & 0.0777 & 0.040 & 1.159 \\
			1.510 & 0.0059 & 0.143 & 0.305 				&& 2.408 & 0.6209 & 0.107 & 24.635 \\
			1.651 & 0.0298 & 0.205 & 2.259 				&& 2.468 & 0.0136 & 0.022 & 0.109 \\
			1.798 & 0.0249 & 0.177 & 1.637 				&& 2.485 & 0.0249 & 0.030 & 0.272 \\
			1.884 & 0.0191 & 0.122 & 0.865 				&& 2.507 & 0.0189 & 0.039 & 0.270 \\
			1.933 & 0.0159 & 0.072 & 0.434 				&& 2.540 & 0.0139 & 0.055 & 0.285 \\
			2.052 & 0.1311 & 0.137 & 6.694 				&& 2.585 & 0.0048 & 0.077 & 0.199 \\
			2.134 & 0.2637 & 0.082 & 8.044 				&& 2.664 & 0.0003 & 0.063 & 0.007 \\
			2.231 & 0.7846 & 0.099 & 28.843 				&& 2.759 & 0.0023 & 0.259 & 0.227 \\
			2.279 & 0.0898 & 0.042 & 1.407 				&& 2.981 & 0.0006 & 0.224 & 0.070 \\
			2.321 & 0.8120 & 0.070 & 20.944 				&&  & &  &  \\
			\hline	
		\end{tabular} 
\end{table}


\clearpage
\appendix 
\newcounter{global}
\newcounter{ESM_fig_number}[global]
\setcounter{ESM_fig_number}{0}
\newcounter{esm}[global] 
\setcounter{esm}{1}

\section{Online Supplementary Material~\arabic{esm}}
\label{ESM1}

\setcounter{ESM_fig_number}{0}

\begin{figure}[th]
	\addtocounter{ESM_fig_number}{1}
	\centering
	\includegraphics[height=.37\textwidth]{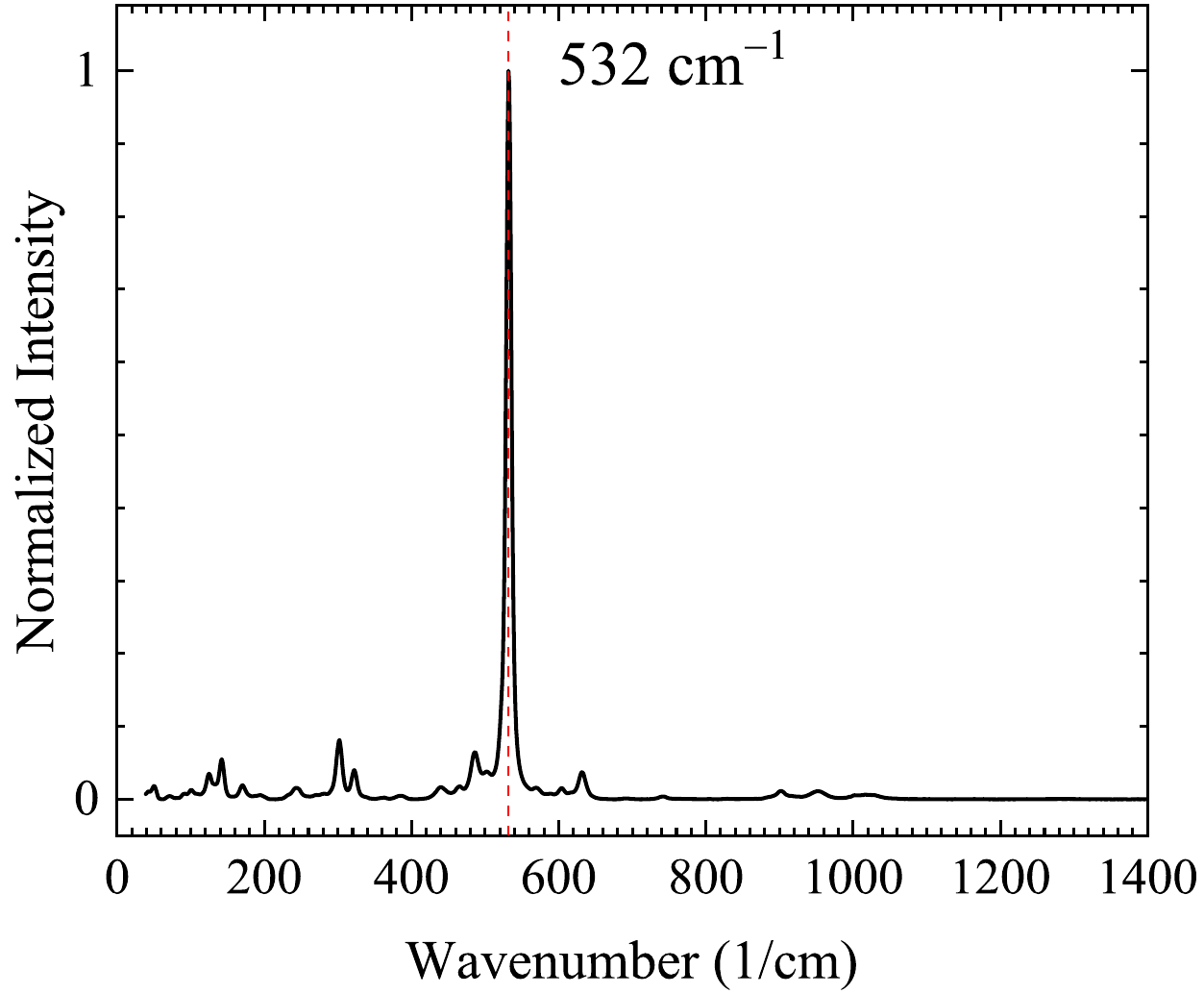}\,(a)
	\includegraphics[height=.37\textwidth]{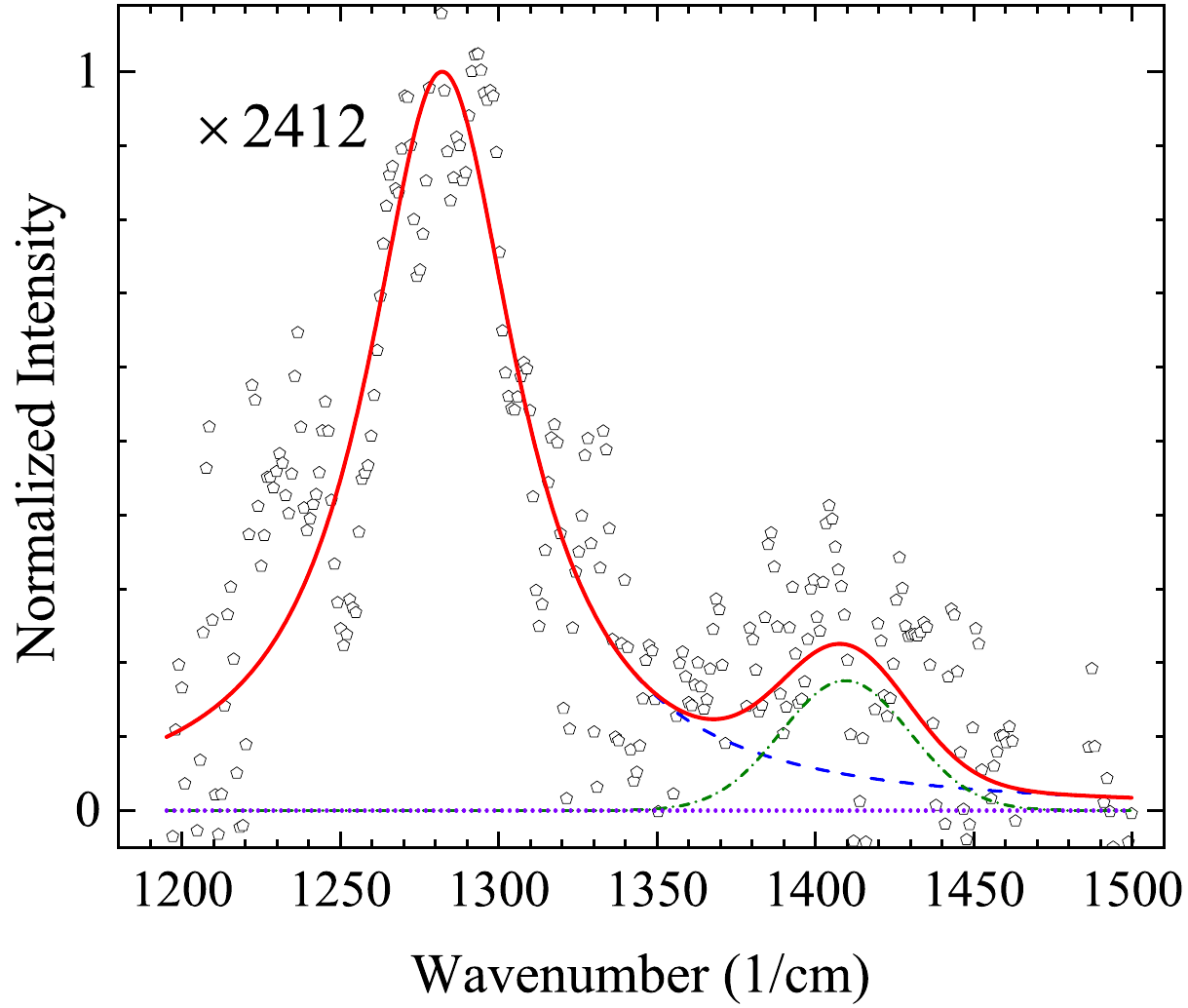}\,(b)\\
		Figure~S\arabic{ESM_fig_number}:
		Survey Raman spectrum of orthorhombic KSbOSiO$_4$ \cite{KSS_Raman_spectrum} (a) and its weak bands detected beyond the wavenumber of 1200\,cm$^{-1}$ (b).
\end{figure}

\end{document}